\documentclass[twocolumn,showpacs,preprintnumbers,superscriptaddress,prl,amsmath,amssymb,nofootinbib]{revtex4-1}

\usepackage{graphicx}
\usepackage{dcolumn}
\usepackage{bm}
\usepackage{color}
\usepackage{ulem}
\usepackage{gensymb}
\usepackage{braket}
\usepackage{epstopdf}
\usepackage{amsmath}
\usepackage[percent]{overpic}

\begin{document}

\title{Magnetic order in single crystals of Na$_3$Co$_2$SbO$_6$ with a honeycomb arrangement of 3d$^7$ Co$^{2+}$ ions}

\author{J.-Q. Yan}
\email{yanj@ornl.gov}
\affiliation{Materials Science and Technology Division, Oak Ridge National Laboratory, Oak Ridge, Tennessee 37831, USA}

\author{S. Okamoto}
\affiliation{Materials Science and Technology Division, Oak Ridge National Laboratory, Oak Ridge, Tennessee 37831, USA}

\author{Y. Wu}
\affiliation{Neutron Scattering Division, Oak Ridge National Laboratory, Oak Ridge, Tennessee 37831, USA}

\author{Q. Zheng}
\affiliation{Materials Science and Technology Division, Oak Ridge National Laboratory, Oak Ridge, Tennessee 37831, USA}

\author{H. D. Zhou}
\affiliation{Department of Physics, University of Tennessee, Knoxville, Tennessee 37996, USA}

\author{H. B. Cao}
\affiliation{Neutron Scattering Division, Oak Ridge National Laboratory, Oak Ridge, Tennessee 37831, USA}

\author{M. A. McGuire}
\affiliation{Materials Science and Technology Division, Oak Ridge National Laboratory, Oak Ridge, Tennessee 37831, USA}

\date{\today}

\begin{abstract}
We have synthesized single crystals of Na$_3$Co$_2$SbO$_6$ and characterized the structure and magnetic order by measuring anisotropic magnetic properties, heat capacity, x-ray and neutron single crystal diffraction. Magnetic properties and specific heat of polycrystalline Na$_3$Co$_2$SbO$_6$ were also measured for comparison. Na$_3$Co$_2$SbO$_6$ crystallizes in a monoclinic structure (space group $C2/m$) with [Co$_2$SbO$_6$]$^{3-}$ layers separated by Na$^+$ ions. The temperature dependence of magnetic susceptibility shows significant anisotropic behavior in the whole temperature range 2\,K-350\,K investigated in this work. An effective moment of about 5.5\,$\mu_B$/Co$^{2+}$ from a Curie-Weiss fitting of the magnetic  susceptibility is larger than the spin only value and signals significant orbital contribution.  Na$_3$Co$_2$SbO$_6$ single crystal undergoes a transition into a long-range antiferromagnetically ordered state below $T_N$=5\,K. Neutron single crystal diffraction confirmed the zigzag magnetic structure with a propagation vector k\,=\,(0.5, 0.5, 0). The ordered moment is found to be 0.9\,$\mu_B$ at 4\,K and align along the crystallographic \textit{b}-axis. Density functional theory calculations suggest that the experimentally observed zigzag order is energetically competing with the Neel order. It is also found that the covalency between Co $d$ and O $p$ is quite strong and competes with the local spin-orbit coupling, suggesting a $J_{eff}$=1/2 ground state may not be realized in this compound.

\end{abstract}

\maketitle

\section{Introduction}
$4d$/$5d$ transition metal compounds such as \textit{A}$_2$IrO$_3$ (\textit{A}=Li, Na) and $\alpha$-RuCl$_3$ have been intensively studied in recent years for possible exotic quantum spin liquid state.\cite{takayama2015hyperhoneycomb,singh2010antiferromagnetic,plumb2014alpha} These compounds have a quasi-two-dimensional structure with weak coupling between layers of honeycomb arranged transition metal ions. The strong spin-orbit coupling in these 4d/5d transition metal ions leads to the effective pseudospin ($J_{eff}=1/2$) electronic state. The in-plane honeycomb arrangement of transition metal ions enables the orbital-dependent hopping processes giving rise to the Kitaev-type interactions. However,
 all these compounds show a long range magnetic order at low temperatures due to the Heisenberg exchange interactions. Therefore, various approaches have been employed to suppress the long range magnetic order, such as high magnetic fields\cite{sears2017phase,zheng2017gapless,baek2017evidence,leahy2017anomalous} and chemical substitution\cite{kitagawa2018spin,lampen2017destabilization,koitzsch2017nearest,do2018short}. Recently, Liu {\it et al}.\citep{liu2018pseudospin} and Sano {\it et al}.\citep{sano2018kitaev} proposed that the Kitaev model can be realized in materials based on $d^7$ ions with an electronic configuration of $t_{2g}^5e_g^2$ (\textit{S}=3/2, \textit{L}=1), which also host the pseudospin-1/2 magnetism. One interesting feature for these $d^7$  candidates is that in the magnetic honeycomb plane the ferromagnetic interactions from $e_g$ spins compensate the antiferromagnetic interactions resulting from $t_{2g}$ electrons. This provides a new approach to tuning the relative magnitude of Heisenburg and Kitaev interactions for the realization of a spin liquid ground state.

\begin{figure} \centering \includegraphics [width = 0.45\textwidth] {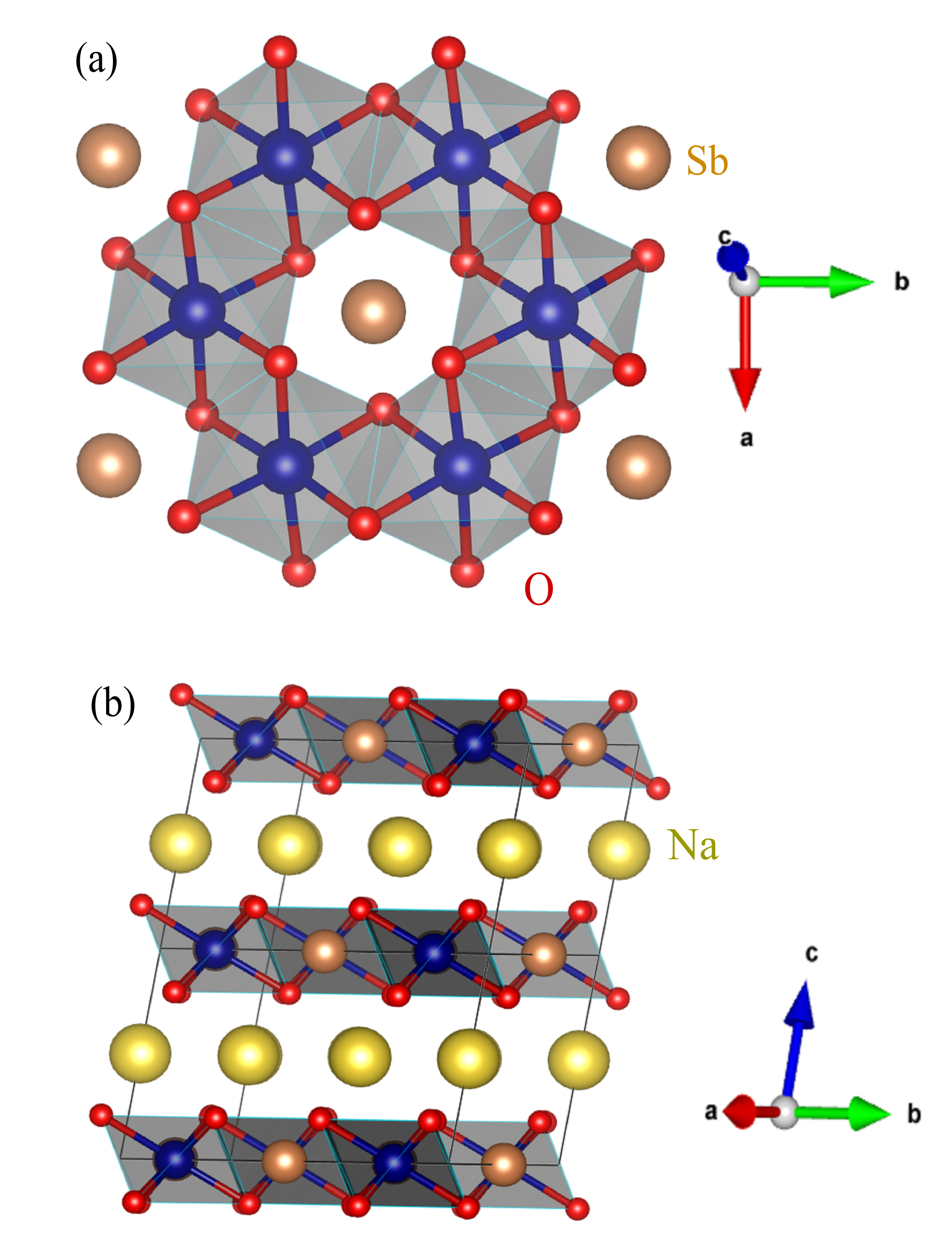}
\caption{(color online) Crystal structure of Na$_3$Co$_2$SbO$_6$ (a) the honeycomb arrangement of Co ions in \textit{ab}-plane, and (b) the honeycomb layers are separated by Na layers.}
\label{crystalStr-1}
\end{figure}

One candidate material proposed by Liu {\it et al}.\citep{liu2018pseudospin} is Na$_3$Co$_2$SbO$_6$.  The monoclinic structure (space group \textit{C}2/\textit{m})\cite{zvereva2016orbitally,wong2016zig} is shown in Fig.\,\ref{crystalStr-1}.  In the magneto-active layer, Na$_3$Co$_2$SbO$_6$ has a honeycomb structure of edge-sharing CoO$_6$ octahedra with one SbO$_6$ octahedron occupying the center of each honeycomb. These layers are separated by Na. Previous studies\cite{wong2016zig,viciu2007structure} on the magnetic properties and specific heat of polycrystalline samples suggest that  Na$_3$Co$_2$SbO$_6$ orders antiferromagnetically below $T_N$=8.3\,K or 4.4\,K which might result from the varying sodium stoichiometry. The Curie-Weiss fitting of the high temperature magnetic susceptibility gives a nearly zero Weiss constant and an effective moment around 5\,$\mu_B$/Co$^{2+}$ larger than the expected spin-only value, indicating a significant orbital contribution. Neutron powder diffraction measurement\cite{wong2016zig} suggests a partially frustrated ‘zigzag’ ordering with a propagation vector k\,=\,(0.5, 0.5, 0) and a magnetic moment of 1.79(4)\,$\mu_B$ at 1.5\,K aligning nearly normal to the plane of the layer along the crystallographic \textit{c}-axis.

Considering the quasi-two-dimensional structure of Na$_3$Co$_2$SbO$_6$, anisotropic physical properties are expected which can be learned from investigating single crystals. To the best of our knowledge, single crystals of  Na$_3$Co$_2$SbO$_6$ have not been available for a detailed study of the intrinsic and anisotropic properties. In this work, we report the growth of millimeter-sized Na$_3$Co$_2$SbO$_6$ single crystals and the characterization of structure and physical properties. Single crystal x-ray diffraction measurements confirm the monoclinic C2/\textit{m} structure and observe diffuse scattering along L due to the stacking faults in the as-grown crystals. Magnetic measurements suggest significant anisotropic behavior in the whole temperature range 2\,-\,350\,K. Na$_3$Co$_2$SbO$_6$ single crystal orders antiferromagnetically below 5\,K. Neutron single crystal diffraction confirms the zigzag magnetic structure with a propagation vector k\,=\,(0.5, 0.5, 0) as reported in a previous neutron powder diffraction study. However, different from that for the polycrystalline samples, the ordered moment in our single crystals is determined to align along the crystallographic \textit{b}-axis. Our density functional theory (DFT) calculations show that the experimentally found zigzag ordering is energetically competing with the N{\'e}el ordering and suggest that strong covalency should be considered when understanding the electronic state of Na$_3$Co$_2$SbO$_6$.

\section{Experimental details}
Polycrystalline Na$_3$Co$_2$SbO$_6$ was synthesized by conventional solid state reaction. The starting materials Na$_2$CO$_3$ (99.995\%, Alfa Aesar), Co$_3$O$_4$ (99.7\%, Alfa Aesar), and Sb$_2$O$_3$ (99.9\%, Alfa Aesar) were mixed in the molar ratio of 1.65:0.67:0.5. A 10\% excess Na$_2$CO$_3$ was added to compensate the evaporation loss during sintering. The homogeneous powder was pelletized and then placed in an alumina crucible with a lid and heated to 800$^\circ$C in 20 hrs and held at this temperature for 48 hrs. The resulting product was ground and pelletized again and heated to 900$^\circ$C and held there for 48 hrs before cooling to room temperature. Room temperature x-ray powder diffraction confirmed that the sample is single phase after the second sintering. The refinement of all powder diffraction patterns in this work is performed using the monoclinic \textit{C}2/\textit{m} structure following ref[\citenum{zvereva2016orbitally,wong2016zig}] and our single crystal x-ray diffraction study. The lattice parameters  from Rietveld refinement are \textit{a}=5.3557(8)~{\AA}, \textit{b}=9.2701(10)~{\AA}, \textit{c}=5.6486(6)~{\AA}, and $\beta$=108.40(2)$^\circ$. The lattice parameters are comparable to those reported by Wong {\it et al}.\cite{wong2016zig}

A dense pellet of Na$_3$Zn$_2$SbO$_6$ was also synthesized using a similar procedure. The temperature dependence of specific heat of Na$_3$Zn$_2$SbO$_6$ was used as the phonon reference when estimating the entropy release across magnetic order of Na$_3$Co$_2$SbO$_6$.

\begin{figure} \centering \includegraphics [width = 0.36\textwidth] {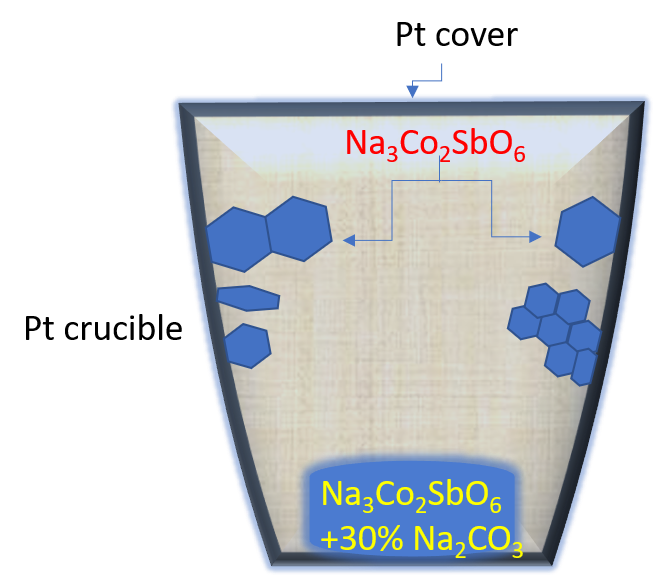}
\caption{(color online) Schematic picture illustrating the growth of Na$_3$Co$_2$SbO$_6$ crystals.}
\label{growth-1}
\end{figure}

To grow single crystals of Na$_3$Co$_2$SbO$_6$, the above fired pellets were reground together with 30wt\% excess Na$_2$CO$_3$. After a thorough mixing, the mixture was pelletized and placed in a Pt crucible with a lid (see Fig.\,\ref{growth-1}). The growth ampoule was then put in a box furnace and heated to 1100$^\circ$C in 10 hrs, held there for 48 hrs, and then cooled to 800$^\circ$C in 24 hrs. The furnace was then shut down to cool to room temperature.

Some crystals were ground into fine powder for x-ray powder diffraction. Room temperature X-ray powder diffraction patterns were collected on a PANalytical X’Pert Pro MPD with a Cu-K$_{\alpha,1}$ incident-beam monochromator. Elemental analysis confirmed the atomic ratio of Na, Co, and Sb, which was carried out on as-grown crystals using a Hitachi-TM3000 microscope equipped with a Bruker Quantax 70 EDS system.

The magnetic properties were measured with a Magnetic Property Measurement System (Quantum Design, QD) in the temperature interval  2 K $\leq$ x $\leq$ 350 K. Specific heat was measured with a QD Physical Property Measurement System (PPMS) in the temperature interval 1.90 K $\leq$ x $\leq$ 250 K. We tried to measure electrical resistivity of Na$_3$Co$_2$SbO$_6$ crystals; however, the total resistance is too large and no reliable resistivity data could be obtained.

X-ray single crystal diffraction was performed to determine the crystal structure using a  Rigaku  single crystal X-ray diffractometer at SNS, ORNL. Neutron single crystal diffraction was performed to determine the magnetic structure at the HB-3A four-circle diffractometer at the High Flux Isotope Reactor of the Oak Ridge National Laboratory. A neutron wavelength of 1.003\,${\AA}$  was used from a bent perfect Si-331 monochromator.\cite{chakoumakos2011four} The neutron diffraction data refinements are based on 85 magnetic and nuclear reflections with the program Fullprof.

\section{Results and discussion}
\subsection{X-ray single crystal diffraction}
\begin{figure} \centering \includegraphics [width = 0.4\textwidth] {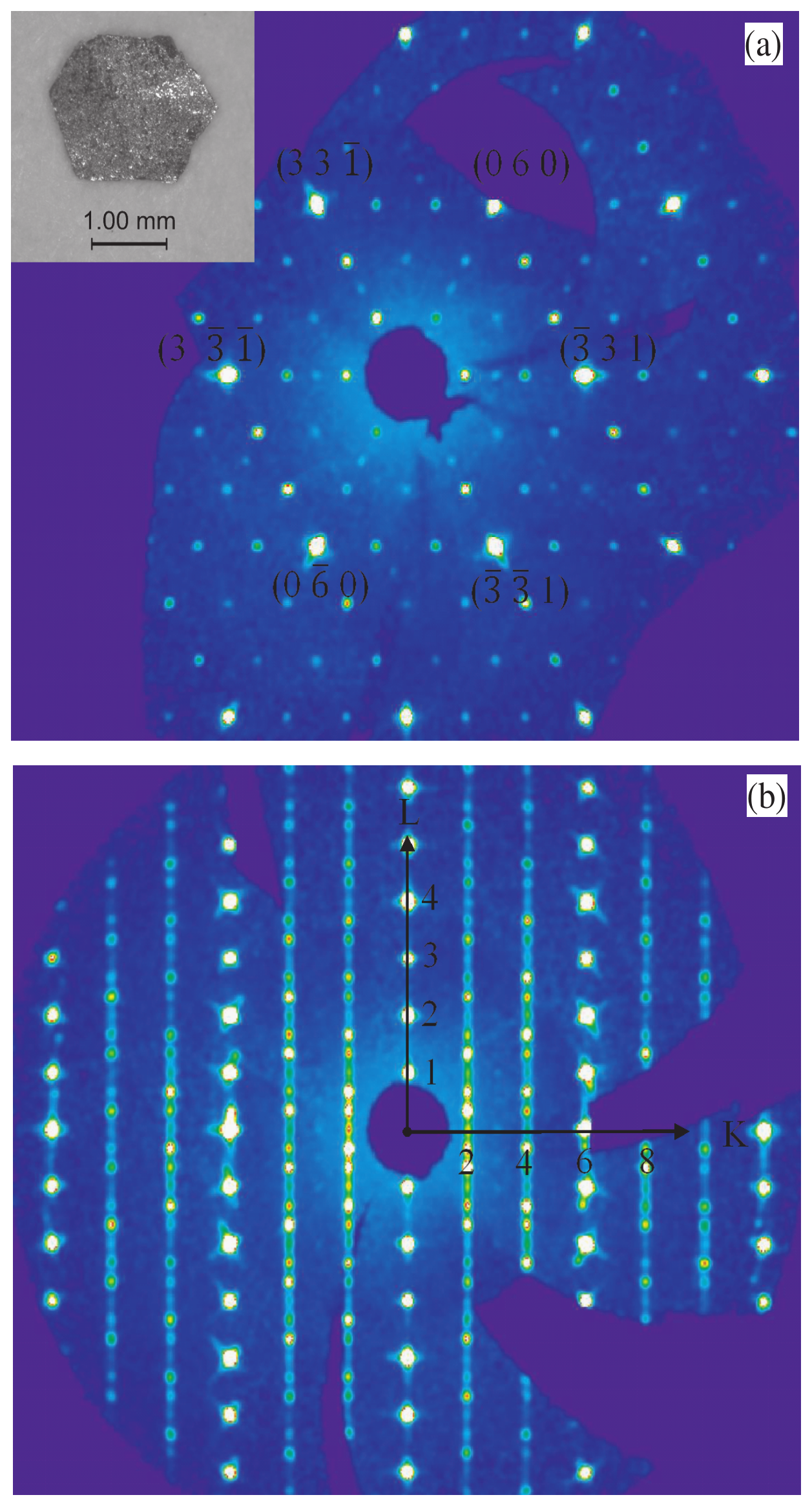}
\caption{(color online) (a) Single crystal X-ray diffraction pattern with the corresponding monoclinic (h k l) indexes.  The reciprocal plane is normal to c*.  (b) The [0 k l ] diffraction cut with index guidance in k and l direction.
Inset of (a) shows one typical crystal with a regular hexagon shape. }
\label{Diffuse-1}
\end{figure}

Plate-like single crystals with a regular hexagon shape were found on the inner surface of the Pt crucible as illustrated in Fig.\,\ref{growth-1}. After crystal growth, the starting pellet was porous and some voids were observed. These observations suggest that the crystal growth occurs with the aid of Na$_2$CO$_3$ vapor. As shown in Fig.\,\ref{Diffuse-1}(a), the as-grown hexagon-shaped crystals are typically 1-2\,mm long for each edge. Room temperature x-ray powder diffraction of pulverized single crystals observed no impurity. Rietveld refinement of the diffraction pattern using monoclinic \textit{C}2/\textit{m} gives the lattice parameters of \textit{a}=5.3708(3)~{\AA}, \textit{b}=9.2631(5)~{\AA}, \textit{c}=5.6485(4)~{\AA}, and $\beta$=108.50(2)$^\circ$. Compared to the polycrystalline samples used for the crystal growth (see experimental details), Na$_3$Co$_2$SbO$_6$ single crystals have a larger \textit{a}-lattice. As discussed later, this \textit{a}-lattice seems to be a good indicator of the magnetic ordering temperature.

To further confirm the monoclinic symmetry of our single crystals, we performed x-ray single crystal diffraction at room temperature. The x-ray diffraction pattern shown in Fig.\,\ref{Diffuse-1}(a))shows well-shaped spots close to trigonal symmetry. Automatic indexing of our single crystal diffraction data indeed tends to return a hexagonal unit cell but relative intensities of the reflections indicated a slight preference for monoclinic symmetry. In the reported crystal structure of Na$_3$Co$_2$SbO$_6$ the individual Na layers and the individual Co$_2$SbO$_6$ slabs have very nearly six-fold symmetry, and it is the manner in which the layers are displaced relative to one another in the stacking sequence that results in an overall monoclinic symmetry.\cite{zvereva2016orbitally} Diffuse spot shapes along the stacking directing indicate stacking faults are prevalent in the crystals, which may be partly responsible for the ambiguity in symmetry. The stacking faults can be modeled approximately by allowing Co and Sb to mix on their respective sites, as described below. We checked all the integrated intensities in trigonal symmetry and also monoclinic symmetry within WinGX. \textit{P}6\textit{/mmm} and \textit{C}2\textit{/m} give comparable merging R-factors. However, a careful investigation of the Bragg peak intensities shown in Fig.\,\ref{Diffuse-1}(b) suggests that the trigonal symmetry is not satisfied. Our analysis concluded space group symmetry is \textit{C}2\textit{/m} which gives the smallest merging R-factor of 0.012.

\begin{table}[htb]
\caption{Selected crystallographic data and refinement parameters for single crystal x-ray diffraction study at room temperature.}
  \begin{tabular}{l p{1.0in} l l} \hline\hline
    Formula & & {\bf Na$_3$Co$_2$SbO$_6$} \\
    Space Group & & C2/m \\
    $a$ (\AA) &  & 5.3707(3)\\
    $b$ (\AA) & & 9.2891(6)\\
    $c$ (\AA) & & 5.6533(4)\\
    $\alpha$ & & 90.00 \\
    $\beta$ & &   108.566(6) \\
    $\gamma$ & & 90.00 \\
    $V$ (\AA$^3$) & & 267.36(3) \\
    $Z$ & & 2 \\
    Crystal size (mm) & & $0.03 \times 0.06 \times 0.08$ \\
    Temperature (K) & & 297 \\
    $\theta$ Range ($^\circ$) & & 4.34 - 32.07 \\
    $\mu$ (mm$^{-1}$) & & 41.773 \\
    $\lambda$ (\AA) & &(Mo K$\alpha$) 0.71073\\
     Collected reflections & & 1713 \\
    Unique reflections & & 444\\
    $R_{int}$ & & 0.0119 \\
    $\sigma I / I$ & & 0.0099 \\
    $h$ & & $-7 \le h \le 7$ \\
    $k$ & & $-13 \le k \le 9$ \\
    $l$ & & $-7 \le l \le 7$ \\
\hline\hline
\end{tabular}
\end{table}

\begin{table*}[!ht]

\caption{\label{str2}The structure parameters of  Na$_3$Co$_2$SbO$_6$ measured at 297 K by single crystal x-ray diffraction. The space group is $C 2/m$. The lattice parameters are $a$=5.3707(3)\AA, $b$=9.2891(6)\AA,
    $c$=5.6533(4)\AA, $\alpha$=90$^o$, $\beta$=108.566(6)$^o$, $\gamma$=90$^o$. $R_{factor}$=1.80 \%. $\chi^2$=1.36. $U$ have units of \AA$^2$. Co/Sb antisite defects are included in the refinement to account for the intensity change from diffuse scattering.}
\begin{tabular}{c|ccc|c|c|c|c|c|c|c}
\hline
atom & $x$& $y$ & $z$ &$Occ.$  & $U_{11}$ & $U_{22}$ &$U_{33}$ &$U_{23}$ &$U_{13}$ &$U_{12}$ \\
\hline
Co1	&0.00000(0)&	0.66635(5)&0.00000(0)&0.922(1)&0.0039(2)&0.0011(1)&0.0055(2)&0&0.0019(2)&	0\\
Sb11	&0.00000(0)&	0.66635(5)&0.00000(0)&0.078(1)&0.0039(2)&0.0011(1)&0.0055(2)&0&0.0019(2)&	0\\
Sb1	&0.00000(0)&0.00000(0)&0.00000(0)&0.844(1)&0.0023(2)&0.0006(1)&0.0055(2)&0&0.0013(2)&	0\\
Co11	&0.00000(0)&0.00000(0)&0.00000(0)&0.156(1)&0.0023(2)&0.0006(1)&0.0055(2)&0&0.0013(2)&	0\\
O1	&0.27299(44)&0.34125(25)&0.79466(48)&	1	&0.0083(9)&0.0025(3)&0.0057(9)&-0.0004(4)&0.0010(7)&-0.0007(4)\\
O2	&0.25106(66)&0.50000(0)&	0.20436(68)&1&0.0098(13)&0.0031(4)&0.0053(12)&	0	&0.0029(10)&	0\\
Na1	&0.00000(0)&0.50000(0)&0.50000(0)&	0.920(3)&0.0066(11)&0.0019(4)&0.0064(11)&	0	&0.0005(8)&	0\\
Na2	&0.50000(0)&	0.32755(21)&0.50000(0)&0.980(4)&0.0075(8)&0.0023(3)&0.0066(8)&	0	&0.0027(6)&	0\\

\hline
\end{tabular}

\end{table*}

A summary of single crystal x-ray diffraction data and refinement parameters for Na$_3$Co$_2$SbO$_6$ crystals is provided
in Table I. Altogether 444 unique peaks were obtained at room temperature with R$_{int}$= 1.2$\%$. The refinement was performed with Fullprof suite and it is well converged with $R_{factor}$=1.80 \% and $\chi^2$=1.36. It is worth mentioning that the refinement was performed with $\sim$10\% Co/Sb antisite disorder, i.e. Co and Sb switch positions.  Including Co/Sb antisite disorder can significantly improve the fitting as this accounts for the peak intensity suppression by the stacking faults. In an octahedral environment, the ionic radius of 0.745~{\AA} for Co$^{2+}$ is larger than 0.60~{\AA} for Sb$^{5+}$.\cite{shannon1976revised} We cannot rule out the possibility of Co/Sb antisite disorder. However, considering the observation of stacking faults and the fact that sliding the honeycomb layers along any honeycomb bond direction leads to Co/Sb on top of Sb/Co sites, we tend to look at the amount of antisite defects as an indicator for the amount of stacking faults. Table II shows the atomic coordinates, occupancy, and anisotropic displacement parameters. The refinement suggests Na sites are not fully occupied. This Na nonstoichiometry might account for the variation of $T_N$ of different Na$_3$Co$_2$SbO$_6$ samples.

The formation of stacking fault may be an intrinsic feature of Na$_3$Co$_2$SbO$_6$ or come from the crystal growth. As suggested by Zvereva {\it et al}.,\cite{zvereva2016orbitally} the monoclinic and the trigonal structures are very similar in both metrics and lattice energy, which leads to the natural formation of stacking faults. On the other hand, the growth technique employed in this work may also facilitate the formation of stacking faults. The observation of Na$_3$Co$_2$SbO$_6$ crystals on the upper part of the Pt crucible signals a vapor transport growth mechanism. Since the Pt crucible is just loosely covered by a Pt lid, all factors affecting the gas flow inside of the Pt crucible and the evaporation of Na$_2$CO$_3$, such as the temperature profile, or the amount of Na$_2$CO$_3$, can affect the growth kinetics and thus formation of lattice defects.

\begin{figure} \centering \includegraphics [width = 0.46\textwidth] {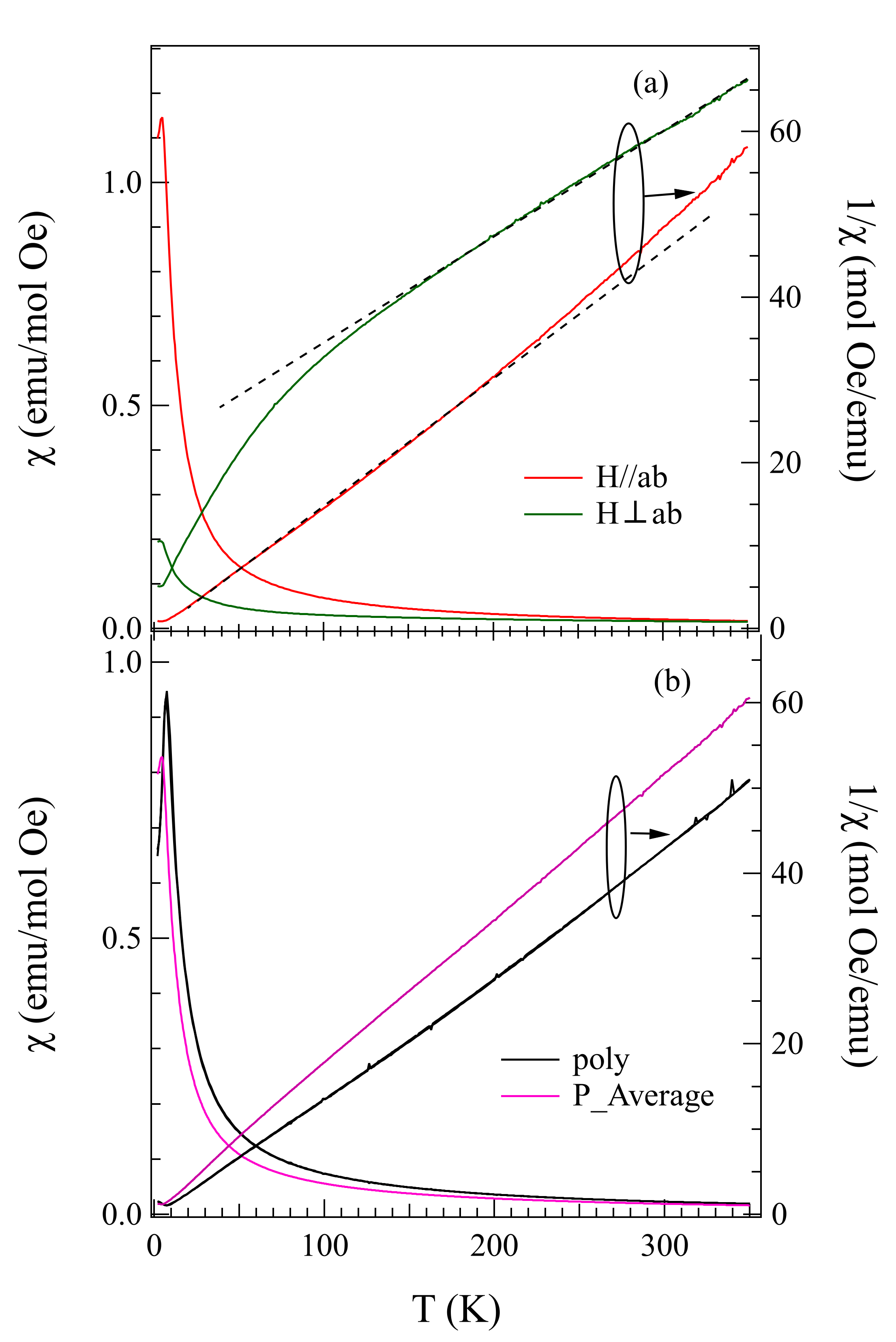}
\caption{(color online) (a) Temperature dependence of magnetic susceptibility, $\chi$(T), and 1/$\chi$(T) measured in a field of 10\,kOe applied parallel to the \textit{ab}-plane and the \textit{c}-axis. The dashed lines highlight the deviation from a linear temperature dependence below $\sim$150\,K for 1/$\chi_\perp$ and above $\sim$200\,K for 1/$\chi_\parallel$.(b) The powder average magnetic susceptibility (P-Average=1/3$\chi_\perp$+2/3$\chi_\parallel$) of single crystals and the temperature dependence of magnetic susceptibility of polycrystalline sample.}
\label{chi-1}
\end{figure}

\subsection{Magnetic properties}
Figure\,\ref{chi-1} shows the temperature dependence of the magnetic susceptibility measured in a magnetic field of 10 kOe applied either perpendicular,$\chi_\perp$, or parallel, $\chi_\parallel$, to the honeycomb plane.  The anomaly around 5\,K in both $\chi_\parallel$ and $\chi_\perp$ curves signals the long range magnetic order, which is further confirmed by specific heat and neutron single crystal diffraction measurements. The magnetic susceptibility is quite anisotropic with $\chi_\parallel$ larger than $\chi_\perp$ in the whole temperature range investigated 2\,-\,350\,K. At room temperature, $\chi_\parallel\approx$1.2$\chi_\perp$. This difference increases with decreasing temperature to $\chi_\parallel\approx$6$\chi_\perp$ around 5\,K. Below 5\,K, $\chi_\perp$ shows little temperature dependence while $\chi_\parallel$ decreases quickly upon further cooling. The different temperature dependence of magnetic susceptibility below 5\,K suggests that the spins are aligned in \textit{ab} plane. This is confirmed by neutron single crystal diffraction measurements presented later.

Figure\,\ref{chi-1}(a) also shows the inverse magnetic susceptibility. The dashed lines highlight the deviation from a linear temperature dependence for 1/$\chi_\perp$ below $\sim$150\,K and for 1/$\chi_\parallel$ above $\sim$250\,K. The latter is similar to that observed for polycrystalline sample.\cite{viciu2007structure} We then fit the $\chi$ data in the temperature range 150\,K$\leq$T$\leq$350\,K using $\chi$(T)=$\chi_0$+C/(T-$\theta$), where $\chi_0$ is the temperature independent term, C is the Curie constant and $\theta$ the Weiss constant. The fitting of $\chi_\perp$ gives $\chi_0$=-2$\times$10$^{-4}$emu/mol, $\theta_c$=-170\,K and $\mu_{eff}$(c)=5.60\,$\mu_B$/Co$^{2+}$. For $\chi_\parallel$, the fitting suggests $\chi_0$=-5$\times$10$^{-3}$emu/mol, $\theta_{ab}$=-4\,K and $\mu_{eff}$(ab)=5.48\,$\mu_B$/Co$^{2+}$, respectively. The large difference between $\theta_{ab}$ and $\theta_{c}$ is surprising. As presented later, our density functional theory calculations suggest these anomalies result from the single ion anisotropy and the antiferromagnetic interlayer coupling in Na$_3$Co$_2$SbO$_6$. The strong single ion anisotropy can lead to abnormal effective Weiss temperatures, which makes the interpretation of the Weiss constant more complicated.\cite{johnston2017influence}

The Co$^{2+}$ ions in an octahedral crystal field have an electronic configuration of t$_{2g}^5e_g^2$. An effective moment of 3.87$\mu_B$/Co$^{2+}$ is expected for S=3/2 (assuming g=2) from the spin-only contribution. The experimental values of $\mu_{eff}$(c)=5.60\,$\mu_B$/Co$^{2+}$ and $\mu_{eff}$(ab)=5.42\,$\mu_B$/Co$^{2+}$ are larger than the expected spin-only value, which signals significant orbital contribution. The above experimental values are also much larger than $\mu_{eff} =3.75 \mu_B$/Co$^{2+}$ expected for the pseudospin-1/2 case proposed theoretically.\citep{liu2018pseudospin,sano2018kitaev}
This theoretical value is determined by $\mu_{eff} = g_J \sqrt{J(J+1)}$ with $g_J = - \frac{3}{2} g_L + \frac{5}{3} g_S$,
where $g_{L ,S}$ are the $g$ factor for the effective orbital moment and the spin moment, respectively, given by $g_L = -\frac{2}{3}$ and $g_S=2$.\citep{khomskii_2014}

The powder average value of magnetic susceptibility was obtained as 1/3$\chi_\perp$+2/3$\chi_\parallel$ and plotted in Fig.\,\ref{chi-1}(b). Also plotted in Fig.\,\ref{chi-1}(b) is $\chi(T)$ of polycrystalline Na$_3$Co$_2$SbO$_6$ used for crystal growth in this study. The powder average is close to $\chi(T)$ of the polycrystalline sample but they do not overlap with some difference especially at low temperatures. The reason is that the polycrystalline sample orders magnetically at $T_N$=7.8\,K where a cusp in $\chi(T)$ curve was observed. This $T_N$ is close to 8.3\,K reported by Wong et al\cite{wong2016zig} for their polycrystalline samples. The magnetic ordering temperature of our single crystals of 4.8\,K is similar to 4.4\,K reported by Viciu {\it et al}.\cite{viciu2007structure} for their polycrystalline samples.

All samples synthesized in different groups have comparable \textit{b}-lattice, \textit{c}-lattice, and $\beta$ angle, but can be classified into two groups following their \textit{a}-lattice: \textit{a}=5.3557(8)~{\AA} of our polycrystalline sample is comparable to \textit{a}=5.3565(1)~{\AA} of Wong's polycrystalline sample; \textit{a}=5.3708(3)~{\AA} for our single crystals is comparable to 5.3681(2)~{\AA} of Viciu's polycrystalline samples. The comparison suggests that samples with a larger \textit{a}-lattice have a lower T$_N$ and \textit{a}-lattice can thus be a good indicator of the Neel temperature.

The variation of $T_N$ can result from Na nonstoichiometry as proposed by Wong {\it et al}.\cite{wong2016zig} Our single crystal diffraction suggests some Na deficiency. It is worth mentioning that the Na content can be purposely controlled by chemical deintercalation.\cite{chou2004thermodynamic,van1980crse2} This might be employed to suppress the long range magnetic order of Na$_3$Co$_2$SbO$_6$. The effects of Na content on the magnetism and structure deserve further investigation. One more factor that might affect T$_N$ is the stacking fault. As evidenced by the x-ray single crystal diffraction measurements, our single crystals have stacking faults. A disturbed stacking is expected to suppress the long range magnetic order of Na$_3$Co$_2$SbO$_6$. However, we want to mention that in $\alpha$-RuCl$_3$ the stacking fault induces magnetic order at temperatures higher than $T_N$=7\,K.\cite{cao2016low}

\begin{figure} \centering \includegraphics [width = 0.46\textwidth] {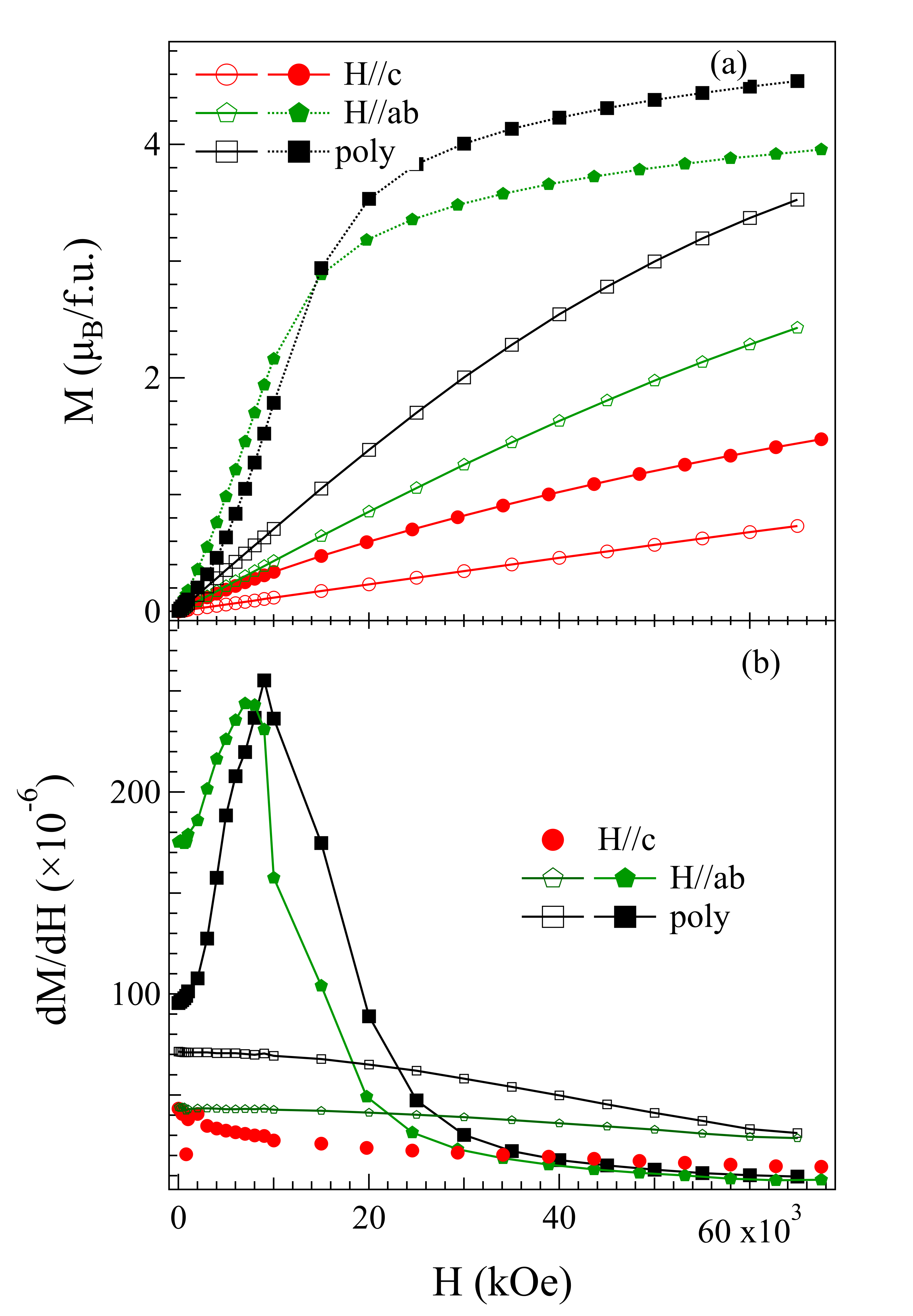}
\caption{(color online) (a) Field dependence of magnetization of single crystals and polycrystals at 2\,K (solid) and 30\,K (open), respectively. (b) Derivative of the magnetization curves at 2\,K (solid) and 30\,K (open).}
\label{MH-1}
\end{figure}

Figure\,\ref{MH-1} shows the field dependence of magnetization for polycrystalline sample and single crystals. While a nearly linear field dependence was observed when the field is applied long \textit{c}-axis, the magnetization of polycrystalline samples and single crystals with field applied in \textit{ab}-plane tends to saturate at high fields even at 30\,K.  At 2\,K, the magnetization in an applied field of 60\,kOe is 2.25\,$\mu_B$/Co$^{2+}$ for our polycrystalline sample, slightly larger than 2\,$\mu_B$/Co$^{2+}$ for our single crystals. Both values are comparable to those reported previously\cite{viciu2007structure,wong2016zig}, however, much larger than 0.70\,$\mu_B$/Co$^{2+}$ when the external magnetic field is applied along the \textit{c}-axis. The field derivative, $dM/dH$, shows a maximum around 10\,kOe for polycrystalline samples and 9\,kOe for single crystals with field applied in \textit{ab}-plane, signaling the field induced spin-flop transition.\cite{viciu2007structure,wong2016zig} However, this feature is absent for single crystals when the magnetic fields are applied along the \textit{c}-axis. This anisotropic behavior suggests that the magnetic moments are aligned in the \textit{ab}-plane. This is consistent with the anisotropic temperature dependence of magnetic susceptibility shown in Fig.\,\ref{chi-1} and our neutron single crystal diffraction measurements further confirm that the moment aligns along the crystallographic \textit{b}-axis.

\begin{figure} \centering \includegraphics [width = 0.46\textwidth] {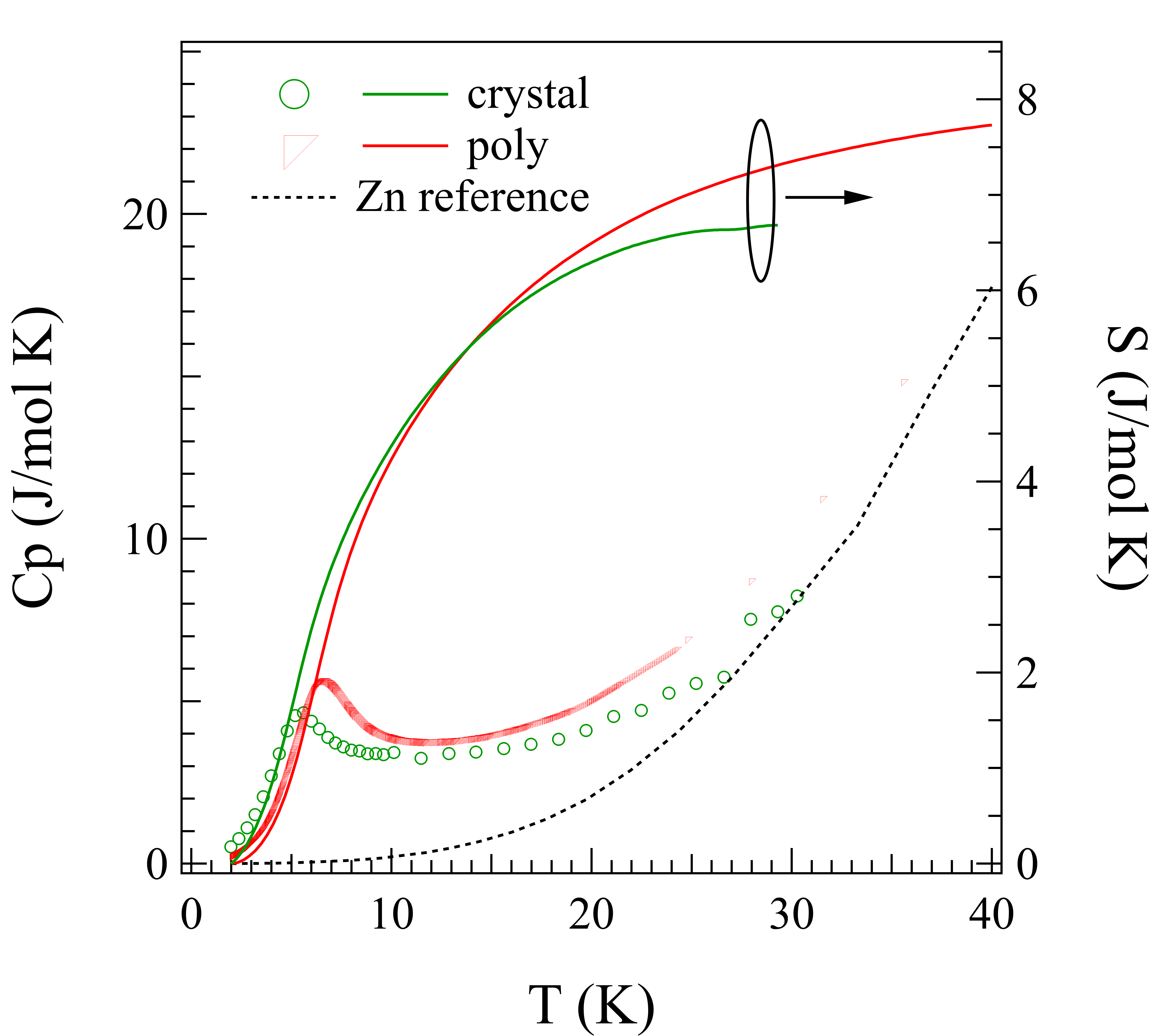}
\caption{(color online) Specific heat and entropy release across the magnetic order of Na$_3$Co$_2$SbO$_6$ single crystal and polycrystals. The dashed curve shows the specific heat of Na$_3$Zn$_2$SbO$_6$ which is used as the phonon reference.}
\label{CpEntropy-1}
\end{figure}

\subsection{Specific heat}
Figure\,\ref{CpEntropy-1} shows the temperature dependence of the specific heat of both polycrystalline pellet and single crystals. A lambda-type anomaly signals the occurrence of a long range magnetic order of Co$^{2+}$ around 5\,K in Na$_3$Co$_2$SbO$_6$ single crystals and around 8\,K for the polycrystalline sample. The magnetic ordering temperature obtained from specific heat data agrees with that determined from magnetic susceptibility shown in  Fig.\,\ref{chi-1} and neutron single crystal diffraction presented later in Fig.\,\ref{neutron-1}. In order to estimate the entropy release across the magnetic order, specific heat of Na$_3$Zn$_2$SbO$_6$ was measured and used as the phonon reference. The entropy release is 6.7J/mol K and 7.7J/mol K for the single crystal and the polycrystalline pellet, respectively. The entropy release is much lower than the theoretical spin-only value of 23\,J/mol K for Co$^{2+}$ but larger than 5.76\,J/mol K expected for the pseudospin-1/2 case.

\begin{figure} \centering \includegraphics [width = 0.46\textwidth] {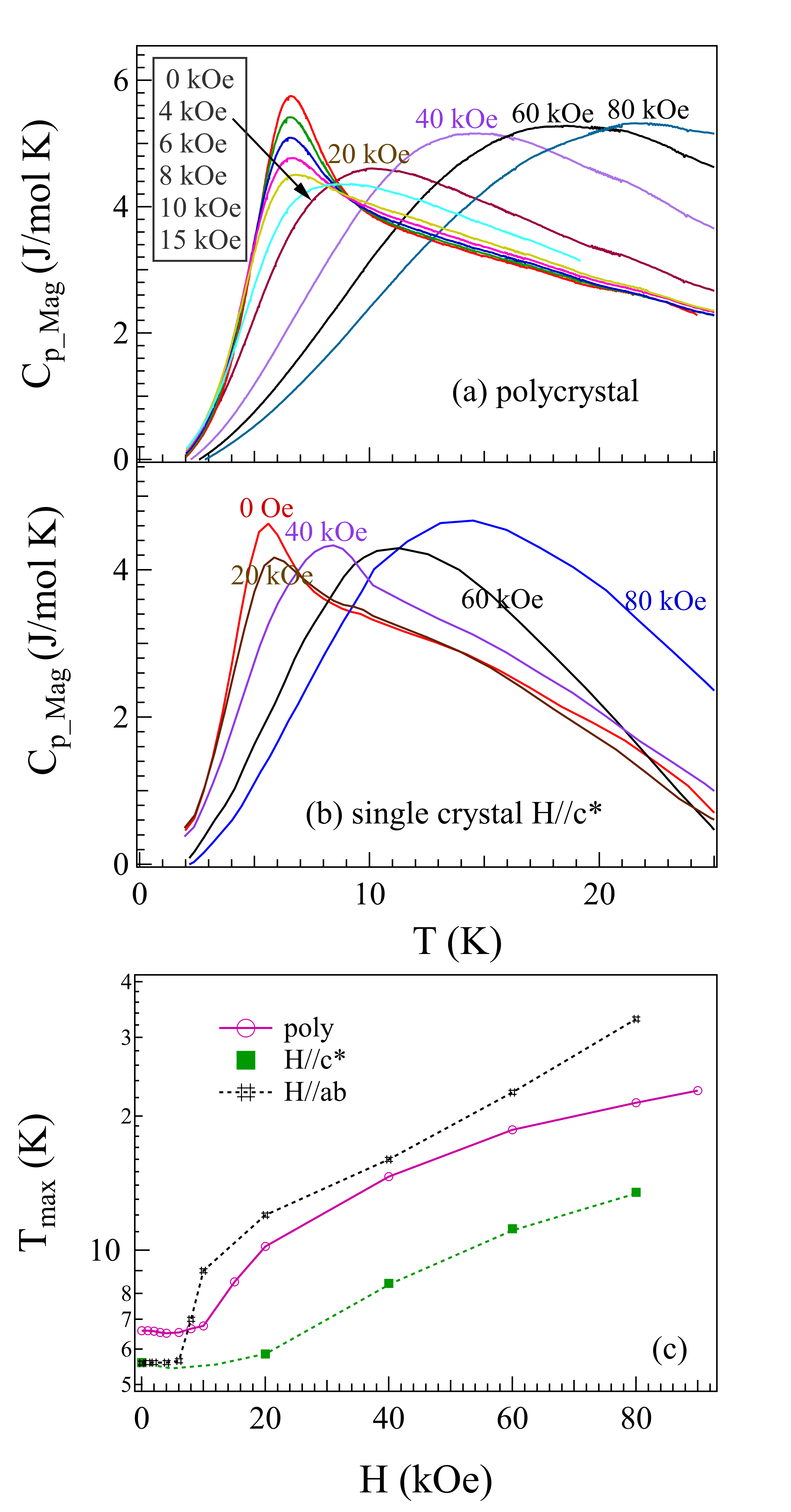}
\caption{(color online) (a, b) Magnetic specific heat in different magnetic fields for (a) polycrystalline and (b) single crystal Na$_3$Co$_2$SbO$_6$ with fields applied perpendicular to the crystal plates. The magnetic specific heat, $C_{P_Mag}$ was obtained by subtracting the specific heat of Na$_3$Zn$_2$SbO$_6$ which is used as phonon reference. (c) Field dependence of $T_{max}$. $T_{max}$ is defined as the temperature where the magnetic specific heat has a maximum. The solid and dashed curves are a guide to the eyes.}
\label{CpH-1}
\end{figure}

Figure\,\ref{CpH-1}(a) and (b) show the field dependence of magnetic specific heat obtained by subtracting the specific heat of Na$_3$Zn$_2$SbO$_6$ from that of Na$_3$Co$_2$SbO$_6$ measured in different magnetic fields. We monitored and plotted in Fig.\,\ref{CpH-1}(c) the field dependence of $T_{max}$ which is defined as the temperature where the magnetic specific heat has a maximum. Below $\sim$6\,kOe, $T_{max}$ of polycrystalline sample decreases from 6.60\,K in zero magnetic field to 6.52\,K in a field of 4\,kOe. Above $\sim$6\,kOe, T$_{max}$ increases rapidly with increasing fields as shown in Figure\,\ref{CpH-1}(c). For the single crystal sample, $T_{max}$ has a larger shift to higher temperatures when the magnetic fields are applied in \textit{ab}-plane than perpendicular to the plane. This field dependence of magnetic order is quite different from that of $\alpha$-RuCl$_3$ where an in-plane field suppresses the long range magnetic order.\cite{sears2017phase,zheng2017gapless,baek2017evidence,leahy2017anomalous}

\subsection{Neutron single crystal diffraction}
\begin{figure} \centering \includegraphics [width = 0.46\textwidth] {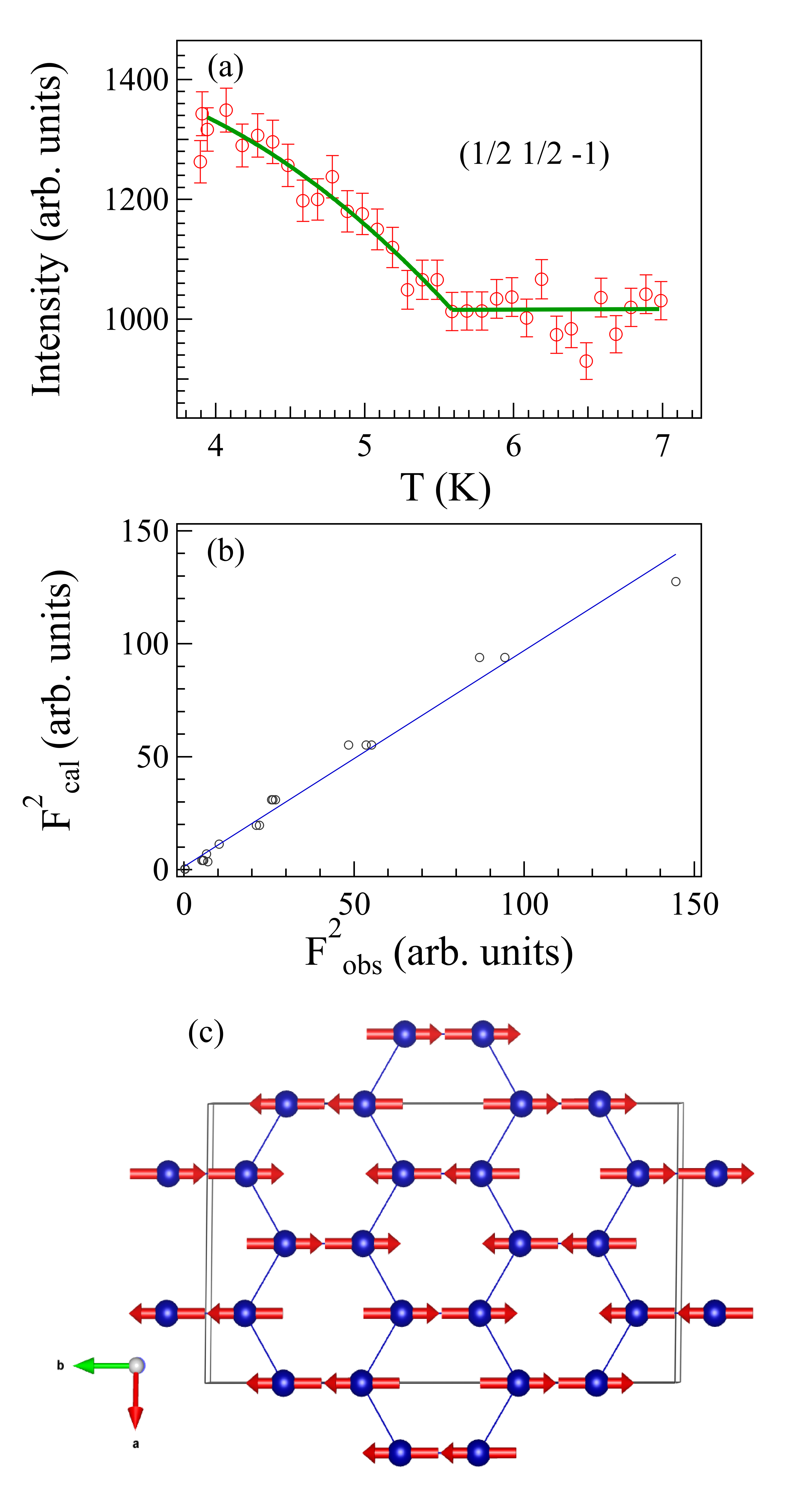}
\caption{(color online) (a) Temperature dependence of the intensity of (0.5 0.5 -1) magnetic reflection. The solid curve is a guide to the eyes. (b)Observed squared structure factors vs the calculated ones from the neutron data refinement. (c) zigzag magnetic arrangement in the magneto-active layer. The magnetic structure has an in-phase arrangement of the zigzag magnetic layers. }
\label{neutron-1}
\end{figure}

To further confirm the magnetic order and to determine the magnetic structure, we performed a neutron single crystal diffraction study using one piece of single crystal of about 1\,mg. Survey of magnetic propagation vectors at 4\,K returns (1/2, 1/2, 0). Figure.\,\ref{neutron-1}(a) shows the evolution with temperature of the intensity of the (0.5 0.5 -1) reflection. The temperature dependence suggests $T_N$=5.5\,K.  The best refinement suggests that the Co moments of 0.90(3) $\mu _ B$/Co at 4\,K are in the honeycomb layers and along the \textit{b}-axis. The fitting quality is displayed in Fig.\,\ref{neutron-1}(b) by plotting the observed squared structure factors versus the calculated ones. The zigzag magnetic arrangement in \textit{ab}-plane is shown in Fig.\,\ref{neutron-1}(c).  The neighboring layers with the same in-plane zigzag arrangement stack on top of each other along the crystallographic \textit{c}-axis.

We noticed that the moment direction along the \textit{c}-axis is proposed based on a neutron powder diffraction study.\cite{wong2016zig} Compared to our single crystals with $T_N$=5\,K, the polycrystalline sample investigated by Wong {\it et al}. has a magnetic order at 8.3\,K and a smaller \textit{a}-lattice. It would be interesting to further investigate the effects of Na content on the magnetic structure.

\subsection{DFT calculations}
In order to understand the magnetic behavior of Na$_3$Co$_2$SbO$_6$,
we performed the density functional theory (DFT) calculations.
We used the generalized gradient  approximation  and  projector  augmented  wave (PAW) approach \cite{Blochl1994}
 as implemented in the Vienna {\it ab initio} simulation package (VASP)\cite{Kresse1999,Kresse1996}.
For Co, Sb and O, standard potentials were used (Co, Sb and O in the VASP distribution), and
for Na a potential in which $p$ states are treated as valence states is used (Na$_{pv}$).
As an energy cutoff, we chose 550~eV.
To account for correlation effects the local $U$ is included on the Co $d$ states \cite{Dudarev1998}.
Here, several values of $U_{eff}=U-J_H$ are considered including $U_{eff}=4.4$~eV \cite{Chen2011}.
The SOC is also turned on.

\begin{figure} \centering \includegraphics [width = 0.46\textwidth] {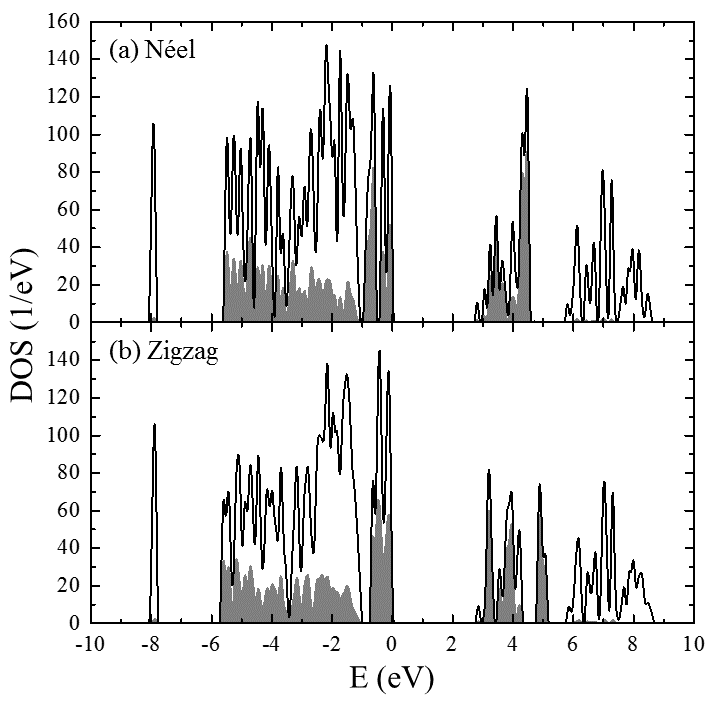}
\caption{Density of states for (a) N{\'e}el ordered state and (b) zigzag ordered state.
Gray areas show the projected DOS on Co $d$ states.The Fermi level is located at $E=0$. Occupied states at $ -1 \alt E<0$~(eV) are mainly from Co $e_g$ majority spin states and $t_{2g}$ minority spin states. Lower energy states located at $-6 \alt E< \alt -1$ are mainly from Co $t_{2g}$ majority spin states and O $p$ states, but Co $t_{2g}$ minority spin states also have significant weights. }
\label{DOS}
\end{figure}

First, we checked the magnetic ground state.
We use the structural parameters obtained in our single crystal x-ray diffraction measurements, but
the unit cell is doubled along $a$ and $b$ directions to accommodate the experimental zigzag magnetic ordering
with the $2\times 2 \times 2$ $k$-point grid.
In addition to the zigzag state, we also considered N{\'e}el ordered state.
With $U_{eff}=4.4$~eV, the N{\'e}el ordered state is found to be more stable than the zigzag state by 0.10~eV per Co.
The relative energy between the two magnetic states is reversed with reducing $U_{eff}$; at $U_{eff}= 1$~eV the zigzag state is more stable than the N{\'e}el state by $0.01$~eV per Co.
Figure~\ref{DOS} shows the total density of state (DOS) and partial DOS projected on Co $d$ states computed with $U_{eff}=4.4$~eV.
The N{\'e}el state (a) and the zigzag state (b) give nearly identical DOSs with an excitation gap $\approx  2.5$~eV. The gap value is reduced to $\approx 0.8$~eV at $U_{eff}=1$~eV.
Thus, the difference between the two magnetic states is rather subtle.

In the presence of trigonal distortion and spin-orbit coupling the ground state doublet is given, in the basis of $|Lz,Sz\rangle$, by:
$|\pm1\rangle$=c1$|\mp1,\pm3/2\rangle$+c2$|0,\pm1/2\rangle$+c3$|\pm1,\mp1/2\rangle$, where the coefficients c1, c2, and c3 are real numbers and are determined by $\delta/\lambda$. $\delta$ describes the trigonal crystal field and $\lambda$ the spin-orbit coupling.\cite{osaki1976anisotropic,lines1963magnetic} Our DFT calculations show that the spin moment $S$ and the orbital moment $L$ are parallel (see table \ref{DFTresults}, $S$ and $L$ have the same sign).
While this is consistent with the $J_{eff}=1/2$ picture as proposed in \citep{liu2018pseudospin,sano2018kitaev},
the ordered spin moment is rather large, comparable to the one expected for the high-spin $S_z=3/2$ state,
and the reduction due to the mixing with $S_z=\pm 1/2$ states seems to be small. This implies that the atomic SOC is competing with the trigonal crystal field in determining the ground state. From our DFT results, we suspect that the covalencey plays an important role to determine the electronic state of Na$_3$Co$_2$SbO$_6$.
As seen in the density of states in Fig.~\ref{DOS},
Co $t_{2g}$ states (mainly majority spins, but minority spins as well) and O $p$ states are strongly hybridized at $-6 \alt E \alt -1$ (eV) below the Fermi level.
Such strong covalency would suppress the mixing between the $S_z=3/2$ state and the $S_z=\pm 1/2$ states by competing with the local SOC.
However, it would be necessary to treat correlation effects more accurately going beyond the DFT framework.

Next, we focus on the on-site spin anisotropy and the interlayer exchange coupling that could provide some insight into the anisotropic magnetic susceptibility.
Those interactions could be modeled by
$H=\sum_{xyzx'y'}(J_\perp^z S_{xyz}^z S_{x'y'z+1}^z + J_\perp^y S_{xyz}^y S_{x'y'z+1}^y) - K \sum_{xyz} (|S_{xyz}^z|^2 -|S_{xyz}^y|^2) +E_0$.
Here, $J_\perp^{z(y)}$ represents the average exchange coupling for the $z(y)$ component of the spins between neighboring two layers, $K$ is the single-site spin anisotropy, and $E_0$ is the energy offset appearing in the DFT calculation.
These parameter values can be estimated by comparing the energy of the following 4 magnetic states:
ferromagnetic states and layered antiferromagnetic states, which have in-plane ferromagnetic spin alignments and out-of-plane antiferromagnetic alignments, with the spin orientation either along the $y$ direction or the $z$ direction.
Considering these magnetic orderings, DFT calculations are carried out using the experimental structure doubled along the $b$ direction and $4 \times 4 \times 2$ $k$-point grid with the other settings mentioned previously.
While the actual magnetic ground state depends on $U_{eff}$,
the sign of interlayer exchange and the single-site anisotropy turned out to be independent of the value of $U_{eff}$;
the out-of-plane exchange is antiferromagnetic, and the anisotopy is uniaxial (table \ref{DFTresults}). 
Note that this does not contradict with the ground state magnetic ordering, either N{\'e}el at large $U_{eff}$ or zigzag at small $U_{eff}$, that has ferromagnetic spin alignment between two neighboring layers.
These magnetic states are mainly stabilized by in-plane magnetic interactions.
While the precise form remains to be determined within the current DFT study,\citep{liu2018pseudospin,sano2018kitaev} in-plane interactions compete with $J_\perp$ and $K$ and are expected to have some frustration arising from either Kitaev and $\Gamma$ terms or longer-range exchange. Thus, the trace of magnetic interactions is expected to become small due to the cancellation between different terms. 

\begin{table}
\newcommand{\lw}[1]{\smash{\lower1.5ex\hbox{#1}}}
\caption{\label{DFTresults}The density functional theory results of the ordered moments, spin $S$ and angular momentum $L$ both in unit of $\mu_B$, out-of-plane exchange $J_\perp^{z,y}$ in meV, and single-site anisotropy $K$ in meV.}
\begin{tabular}{c | c c c c c c c c}
\hline
 \lw{$U_{eff}$ (eV)} & \multicolumn{2}{c}{N{\'e}el} & \multicolumn{2}{c}{zigzag } & \lw{$J_\perp^z S^2$} & \lw{$J_\perp^y S^2$} & \lw{$K S^2$} \\
 & $2S$ & $L$ & $2S$ & $L$ & & &   \\
\hline
4.4 & 2.72 & 0.27 & 2.69 & 0.23 & 0.1 & 0.11 & 3.5 \\
1.0 & 2.57 & 0.26 & 2.58 & 0.19 & 0.23 & 0.16 & 40.2 \\
\hline
\end{tabular}
\end{table}

\section{Summary}

In summary, we have reported the crystal growth, structure, and magnetic order of Na$_3$Co$_2$SbO$_6$ single crystals. Single crystal x-ray diffraction measurements confirm the monoclinic C2/\textit{m} structure and find diffuse scattering along L due to the stacking faults in the as-grown crystals. Magnetic measurements suggest significant anisotropic behavior in the whole temperature range 2\,K-350\,K.  Na$_3$Co$_2$SbO$_6$ single crystal orders magnetically below 5\,K into the zigzag magnetic structure with a propagation vector k\,=\,(0.5, 0.5, 0). The ordered moment is found to be 0.9\,$\mu_B$ at 4\,K and align along the crystallographic \textit{b}-axis. Our DFT calculations showed that the experimentally found zigzag ordering is energetically competing with the N{\'e}el ordering. The zigzag ordering is only stable at relatively small $U_{eff}$ ($\alt 1$~eV). The spin moment and the orbital moment on a Co ion are found to be parallel irrespective of the magnetic ordering.
While this is consistent with the proposed $J_{eff}=1/2$ picture, the ordered spin moment is comparable to the one from a $S_z=3/2$ state, and the influence of the SOC appears to be small. This might indicate that the covalency plays an important role to determine the electronic state of Na$_3$Co$_2$SbO$_6$. Thus, it is necessary to go beyond the standard DFT approach to describe the electronic property of Na$_3$Co$_2$SbO$_6$.

\section{Acknowledgments}
The authors thank A. Banerjee, A. F. May, and B. C. Sales for discussions. Work at ORNL was supported by the U.S. Department of Energy, Office of Science, Basic Energy Sciences, Division of Materials Sciences and Engineering (JQY, QZ, and MAM)
and by the Scientific Discovery through Advanced Computing (SciDAC) program funded by the U.S. Department of Energy, Office of Science, Advanced Scientific Computing Research and Basic Energy Sciences, Division of Materials Sciences and Engineering (SO).
The x-ray and neutron single crystal diffraction work at ORNL was sponsored by the Scientific User Facilities Division, Office of Basic Energy Sciences, U. S. Department of Energy(YW and HC). HDZ thanks the support from NSF-DMR-1350002.

 This manuscript has been authored by UT-Battelle, LLC, under Contract No.
DE-AC0500OR22725 with the U.S. Department of Energy. The United States
Government retains and the publisher, by accepting the article for publication,
acknowledges that the United States Government retains a non-exclusive, paid-up,
irrevocable, world-wide license to publish or reproduce the published form of this
manuscript, or allow others to do so, for the United States Government purposes.
The Department of Energy will provide public access to these results of federally
sponsored research in accordance with the DOE Public Access Plan (http://energy.gov/
downloads/doe-public-access-plan).

\section{references}
%

\end{document}